\documentclass[prl,letter,twocolumn,superscriptaddress,floatfix,nofootinbib,showpacs,amsmath,amssymb,altaffilletter,floatfix]{revtex4-1}

\newcommand{\be}{\begin{equation}}
\newcommand{\ee}{\end{equation}}
\newcommand{\bea}{\begin{eqnarray}}
\newcommand{\eea}{\end{eqnarray}}

\newcommand{\fion}{f_{\rm ion}^{nl}(k',q)}
\newcommand{\fsq}{|f_{\rm ion}^{nl}(k',q)|^2}
\newcommand{\FDM}{$F_{\rm DM}(q)$}
\newcommand{\FDMsq}{|F_{\rm DM}(q)|^2}
\newcommand{\sigmabar}{\overline{\sigma}_e}
\newcommand{\vmin}{v_{\rm min}}
\newcommand{\Eer}{E_{\rm er}}
\newcommand{\Zeff}{Z_{\rm eff}}

\input{ds.def}
\definecolor{ocre}{RGB}{243,102,25}
\definecolor{citrine}{rgb}{0.89, 0.82, 0.04}

\newcommand{\APC}{APC, Universit\'e Paris Diderot, CNRS/IN2P3, CEA/Irfu, Obs de Paris, USPC, Paris 75205, France}
\newcommand{\AQLNGS}{INFN Laboratori Nazionali del Gran Sasso, Assergi (AQ) 67100, Italy}
\newcommand{\AQGSSI}{Gran Sasso Science Institute, L'Aquila 67100, Italy}

\newcommand{\Augustana}{Physics Department, Augustana University, Sioux Falls, SD 57197, USA}
\newcommand{\Belgorod}{Radiation Physics Laboratory, Belgorod National Research University, Belgorod 308007, Russia}
\newcommand{\BHSU}{School of Natural Sciences, Black Hills State University, Spearfish, SD 57799, USA}

\newcommand{\BNL}{Brookhaven National Laboratory, Upton, NY 11973, USA}
\newcommand{\BOINFN}{INFN Bologna, Bologna 40126, Italy}
\newcommand{\BOUniPHY}{Physics Department, Universit\`a degli Studi di Bologna, Bologna 40126, Italy}

\newcommand{\CAUniPHY}{Physics Department, Universit\`a degli Studi di Cagliari, Cagliari 09042, Italy}
\newcommand{\CAINFN}{INFN Cagliari, Cagliari 09042, Italy}

\newcommand{\Campinas}{Physics Institute, Universidade Estadual de Campinas, Campinas 13083, Brazil}

\newcommand{\CIEMAT}{CIEMAT, Centro de Investigaciones Energ\'eticas, Medioambientales y Tecnol\'ogicas, Madrid 28040, Spain}

\newcommand{\CTLNS}{INFN Laboratori Nazionali del Sud, Catania 95123, Italy}
\newcommand{\ENUniCEE}{Engineering and Architecture Faculty, Universit\`a di Enna Kore, Enna 94100, Italy}

\newcommand{\FNAL}{Fermi National Accelerator Laboratory, Batavia, IL 60510, USA}

\newcommand{\GEUni}{Physics Department, Universit\`a degli Studi di Genova, Genova 16146, Italy}
\newcommand{\GEINFN}{INFN Genova, Genova 16146, Italy}

\newcommand{\Hawaii}{Department of Physics and Astronomy, University of Hawai'i, Honolulu, HI 96822, USA}
\newcommand{\Houston}{Department of Physics, University of Houston, Houston, TX 77204, USA}
\newcommand{\IHEP}{Institute of High Energy Physics, Beijing 100049, China}

\newcommand{\INSTM}{Interuniversity Consortium for Science and Technology of Materials, Firenze 50121, Italy}

\newcommand{\JINR}{Joint Institute for Nuclear Research, Dubna 141980, Russia}
\newcommand{\Krakow}{M. Smoluchowski Institute of Physics, Jagiellonian University, 30-348 Krakow, Poland}
\newcommand{\Kurchatov}{National Research Centre Kurchatov Institute, Moscow 123182, Russia}

\newcommand{\Lodz}{Institute of Applied Radiation Chemistry, Lodz University of Technology, 93-590 Lodz, Poland}
\newcommand{\LPNHE}{LPNHE, CNRS/IN2P3, Sorbonne Universit\'e, Universit\'e Paris Diderot, Paris 75252, France}

\newcommand{\MEPhI}{National Research Nuclear University MEPhI, Moscow 115409, Russia}

\newcommand{\MIINFN}{INFN Milano, Milano 20133, Italy}

\newcommand{\MIUni}{Physics Department, Universit\`a degli Studi di Milano, Milano 20133, Italy}
\newcommand{\MSU}{Skobeltsyn Institute of Nuclear Physics, Lomonosov Moscow State University, Moscow 119234, Russia}
\newcommand{\NAINFN}{INFN Napoli, Napoli 80126, Italy}
\newcommand{\NAUniPHY}{Physics Department, Universit\`a degli Studi ``Federico II'' di Napoli, Napoli 80126, Italy}

\newcommand{\Petersburg}{Saint Petersburg Nuclear Physics Institute, Gatchina 188350, Russia}
\newcommand{\PGUniCBB}{Chemistry, Biology and Biotechnology Department, Universit\`a degli Studi di Perugia, Perugia 06123, Italy}
\newcommand{\PGINFN}{INFN Perugia, Perugia 06123, Italy}
\newcommand{\PIINFN}{INFN Pisa, Pisa 56127, Italy}
\newcommand{\PIUniPHY}{Physics Department, Universit\`a degli Studi di Pisa, Pisa 56127, Italy}
\newcommand{\PNNL}{Pacific Northwest National Laboratory, Richland, WA 99352, USA}
\newcommand{\Princeton}{Physics Department, Princeton University, Princeton, NJ 08544, USA}

\newcommand{\RMTreINFN}{INFN Roma Tre, Roma 00146, Italy}
\newcommand{\RMTreUni}{Mathematics and Physics Department, Universit\`a degli Studi Roma Tre, Roma 00146, Italy}
\newcommand{\RMUnoINFN}{INFN Sezione di Roma, Roma 00185, Italy}
\newcommand{\RMUnoUni}{Physics Department, Sapienza Universit\`a di Roma, Roma 00185, Italy}

\newcommand{\SSUniCHP}{Chemistry and Pharmacy Department, Universit\`a degli Studi di Sassari, Sassari 07100, Italy}

\newcommand{\Temple}{Physics Department, Temple University, Philadelphia, PA 19122, USA}

\newcommand{\UCDavis}{Department of Physics, University of California, Davis, CA 95616, USA}
\newcommand{\UCLA}{Physics and Astronomy Department, University of California, Los Angeles, CA 90095, USA}
\newcommand{\UMass}{Amherst Center for Fundamental Interactions and Physics Department, University of Massachusetts, Amherst, MA 01003, USA}

\newcommand{\USP}{Instituto de F\'isica, Universidade de S\~ao Paulo, S\~ao Paulo 05508-090, Brazil}
\newcommand{\VTech}{Virginia Tech, Blacksburg, VA 24061, USA}
\begin{document}
\title{Constraints on Sub-GeV Dark Matter-Electron Scattering from the \DSf\ Experiment}
\author{P.~Agnes}\affiliation{\Houston}
\author{I.~F.~M.~Albuquerque}\affiliation{\USP}
\author{T.~Alexander}\affiliation{\PNNL}
\author{A.~K.~Alton}\affiliation{\Augustana}
\author{G.~R.~Araujo}\affiliation{\USP}
\author{D.~M.~Asner}\affiliation{\BNL}
\author{M.~Ave}\affiliation{\USP}
\author{H.~O.~Back}\affiliation{\PNNL}
\author{B.~Baldin}\altaffiliation{Present address: Raleigh, NC 27613-3133, USA}\affiliation{\FNAL}
\author{G.~Batignani}\affiliation{\PIINFN}\affiliation{\PIUniPHY}
% ...

\collaboration{The \DS\ Collaboration}\noaffiliation
\date{\today}

\begin{abstract}
We present new constraints on sub-GeV dark matter particles scattering off electrons based on 6780.0~kg~d of data collected with the \DSf\ dual-phase argon time projection chamber. This analysis uses electroluminescence signals due to ionized electrons extracted from the liquid argon target. The detector has a very high trigger probability for these signals, allowing for an analysis threshold of three extracted electrons, or approximately 0.05~keVee. We calculate the expected recoil spectra for dark matter-electron scattering in argon and, under the assumption of momentum-independent scattering, improve upon existing limits from XENON10 for dark-matter particles with masses between 30 and 100 MeV/$c^2$.
\end{abstract}


\maketitle

The nature of dark matter (DM) remains unknown despite several decades of increasingly compelling gravitational evidence \cite{Faber:1979em, Refregier:2003jl,Clowe:2006hr,Thompson:2015fm, Ade:2016bk}. While the most favored candidate in a particle physics interpretation is the weakly interacting massive particle (WIMP) \cite{Steigman:1985fk,Bertone:2005bi}, which obtains its relic abundance by thermal freeze-out through weak interactions, there is as yet no unambiguous evidence of WIMP direct detection, therefore warranting searches for other possible DM paradigms.

Another well-motivated class of DM candidates is sub-GeV particles interacting through a vector mediator with couplings smaller than the weak scale. These light DM candidates arise in a variety of models~\cite{Boehm:2004fx,Strassler:2007dx,Hooper:2008hv,Pospelov:2008di,Feng:2008cz}, and there are a number of proposed mechanisms that naturally obtain the expected relic abundance for light DM~\cite{Kaplan:2009ag,Hall:2009bx,Petraki:2013du,Zurek:2013wia,Hochberg:2014dra,Hochberg:2015fj,DAgnolo:2015hf,Harigaya:2016rwr,Kuflik:2016gb,Pappadopulo:2016pkp,Dror:2016rxc,Kuflik:2017dk,DAgnolo:2017dbv,Bernal:2017gd,Berlin:2017ftj}. Light DM may have couplings to electrons, and because the energy transferred by the DM particle to the target depends on the reduced mass of the system, electron targets more efficiently absorb the kinetic energy of sub-GeV-scale light DM than a nuclear target~\cite{Essig:2012iv}. 

There is currently a substantial experimental effort to search for light DM through multiple techniques; see Refs.~\cite{Alexander:2016aln,Battaglieri:2017tk} and references therein. In particular, dual-phase time projection chambers (TPCs) are an excellent probe of light DM, which can ionize atoms to create an electroluminescence signal (S2) even when the corresponding prompt scintillation signal (S1), typically used to identify nuclear recoils, is below the detector threshold~\cite{Essig:2012it}. In this Letter, we present the first limits on light DM-electron scattering from the DarkSide-50 experiment (DS-50). This analysis closely follows Ref.~\cite{Agnes:2018vi}, which contains additional details about the detector, data selection, detector response, and cut efficiencies.

DS-50 is a dual-phase time projection chamber with a \DSfActiveMass\ target of low-radioactivity underground argon (\UAr)~\cite{Agnes:2015gu,AcostaKane:2008im,Back:2012vo,Xu:2015do} outfitted with 38 three-inch PMTs, 19 above the anode and 19 below the cathode. Particle interactions within the target volume create primary UAr scintillation (S1) and ionized electrons. These electrons are drifted towards the anode of the TPC and extracted into a gas layer where they create gas-proportional scintillation (S2). The electron extraction efficiency is better than 99.9\%~\cite{Bondar:2009gh}.  While the trigger efficiency for S1 signals drops to zero below approximately 0.6~keVee, the S2 trigger efficiency remains 100\% above 0.05~keVee due to the high S2 photon yield per electron, \SI{23(1)}{\pe\per\el} in the central PMT as measured by single-electron events caused by impurities within the argon that trap and release single charges. S2 signals are identified off-line using a software pulse-finding algorithm that is effectively 100\% efficient above 0.05~keVee, and a set of basic cuts are applied to the data to reject spurious events. A fiducial cut is then applied that only accepts events whose maximum signal occurs within one of the central seven PMTs in the top PMT array. After all cuts, the detector acceptance is (\num{43(1)})\%, due almost entirely to fiducialization. A correction is applied to events that occur under the six PMTs surrounding the central one to correct for a radial variation in photon yield observed in $^{83m}$Kr source data. 

A DM particle may scatter off a bound electron within the DS-50 detector, ionizing an argon atom. We evaluate the dark-matter recoil spectra for argon following the calculation of Refs.~\cite{Essig:2012iv,Essig:2017bi}. The velocity-averaged differential ionization cross section for bound electrons in the ($n,l$) shell is given by
\bea
&&\frac{d\langle\sigma^{nl}_{\text{ion}} v\rangle}{d \ln \Eer} =\frac{\sigmabar}{8\,\mu_{\chi e}^2}\nonumber\\
&&\times\int dq \, q \, \fsq\, \FDMsq\, \eta(\vmin),
\label{eq:Rdef2}
\eea
where the reference cross section, $\sigmabar$, parametrizes the strength of the interaction and is equivalent to the cross section for elastic scattering on free electrons; $\mu_{\chi e}$ is the DM-electron reduced mass; $q$ is the 3-momentum transfer; $\fion$ is the ionization form-factor, which models the effects of the bound-electron initial state and the outgoing final state perturbed by the potential of the ion from which the electron escaped; $k'$ is the electron recoil momentum; \FDM\ is the DM form factor; and the DM velocity profile is encoded in the inverse mean speed function, $\eta(\vmin) = \langle \frac 1 v \Theta(v - v_{min}) \rangle$, where $\vmin$ is the minimum velocity required to eject an electron with kinetic energy $E_{\text{er}}$ given the momentum transfer $q$ and $\Theta$ is the Heaviside step function.

The details of the argon atom's electronic structure and the outgoing state of the recoil electron are contained in $\fion$, which is a property of the argon target and independent of the DM physics. Computing $\fion$ requires one to model both the initial bound states and the final continuum outgoing states of the electron. The target electrons are modeled as single-particle states of an isolated argon atom described by the Roothaan-Hartree-Fock wave functions. This conservatively neglects the band structure of liquid argon which, if included, should enhance the total electron yield due to the decreased ionization energy in the liquid state~\cite{Kubota:1976cf}. The recoil electron is modeled as the full positive-energy wave function obtained by solving the Schr\"{o}dinger equation with a hydrogenic potential of some effective screened charge $\Zeff$~\cite{Bethe:1977gr}.   We choose a $\Zeff$ that reproduces the energy levels of the argon atom assuming a pure Coulomb potential.  Further details on the computation of $\fion$ are provided in the Appendix.

The DM form factor, \FDM, parametrizes the fundamental momentum-transfer dependence of the DM-electron interaction and has the following limiting values:
\be
F_{\rm DM}(q) = \frac{{m_{A^\prime}}^2 + \alpha^2 {m_e}^2}{{m_{A^\prime}}^2 + q^2} \simeq \begin{cases} 1, \qquad & m_{A^\prime} \gg \alpha m_e \\
\frac{{\alpha}^2 {m_e}^2}{q^2}, \qquad & m_{A^\prime} \ll \alpha m_e \, ,
\end{cases}
\label{eq:FDMdef}
\ee
where $m_{A^\prime}$ is the mass of the vector mediator, $m_e$ is the electron mass, and $\alpha$ is the fine-structure constant. Because \FDM\ is dimensionless by definition, the form factor needs to be defined with respect to a reference momentum scale. The conventional choice is $q_0 = \alpha m_e = 1/a_0$, where $a_0$ is the Bohr radius, because this is typical of atomic momenta. The case where \FDM $\,=1$ corresponds to the ``heavy mediator'' regime, where $m_{A^\prime}$ is much larger than the typical momentum scale. The case where \FDM $\,\propto 1/q^2$ corresponds to the ``light mediator'' regime. 

The inverse mean speed, $\eta(\vmin)$, is defined through the DM velocity distribution in the same way as for GeV-scale WIMPs and nuclear scattering.  We have assumed the standard halo model with escape velocity $v_\text{esc} = 544$~km/s~\cite{Smith:2007fi}, circular velocity $v_0=220$~km/s, and the Earth velocity as specified in~\cite{Lee:2013hg} and evaluated at $t=199$~days ($v_E \approx 244$~km/s), the median run live time for DarkSide-50. Note that the definition of $\vmin$ is different for electron scattering from a bound initial state than for elastic nuclear recoils. The relation $E_{R} = q^2/2m_N$, which is valid in two-body elastic scattering, no longer holds. For a bound electron with principal quantum number $n$ and angular momentum quantum number $l$~\cite{Essig:2017bi}
\be
\vmin (q, E_b^{nl}, \Eer) = \frac{|E_b^{nl}| + \Eer}{q} + \frac{q}{2 m_\chi},
\label{eq:vmindef}
\ee
where $|E_b^{nl}| + \Eer$ is the total energy transferred to the ionized electron, which is a sum of the energy needed to overcome the binding energy, $E_b^{nl}$, and the recoil energy of the outgoing electron, $\Eer$.

\begin{figure}[]
\begin{center}
\includegraphics[width=\columnwidth]{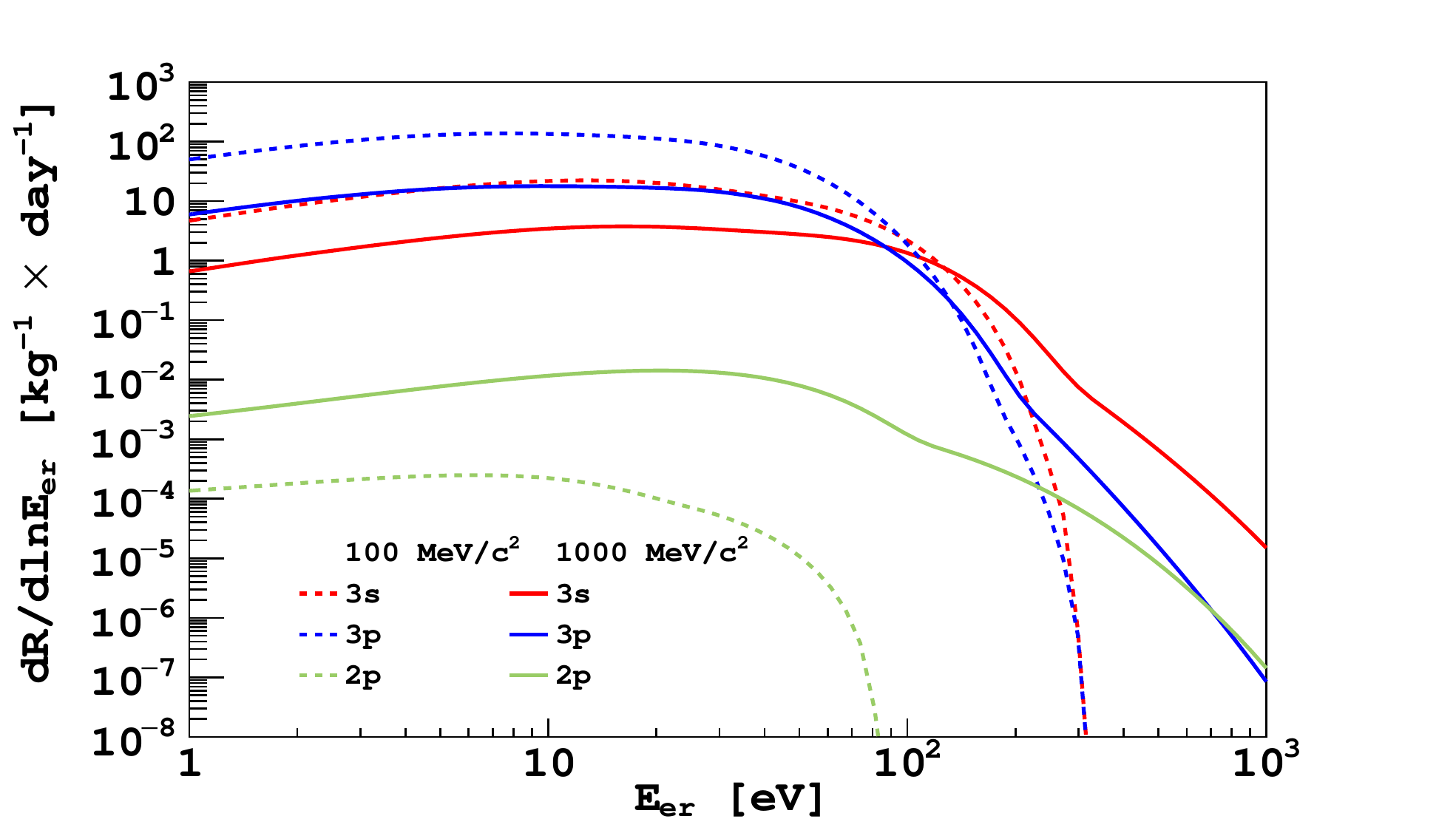}
\end{center}
\caption{\label{fig:ErDMshells} Contributions of the $3s$, $3p$, and $2p$ shells to the DM-electron scattering rate assuming a WIMP-electron cross section of $10^{-36}\ \mathrm{cm}^2$ and $F_{DM}=1$ for a 100~MeV/$c^2$ DM particle (dashed) and a 1000~MeV/$c^2$ DM particle (solid).}
\end{figure}

The velocity averaged differential ionization cross section, Eq.~\ref{eq:Rdef2}, is used to calculate the DM-electron differential ionization rate, 
\be
\frac{dR}{d \ln \Eer} = N_T \frac{\rho_\chi}{m_\chi} \sum_{nl} \frac{d\langle\sigma^{nl}_{\text{ion}} v\rangle}{d \ln \Eer},
\label{eq:Rdef}
\ee 
where $N_T$ is the number of target atoms per unit mass, $\rho_\chi =0.4$~GeV/cm$^3$ is the local DM density used in Ref.~\cite{Essig:2017bi}, and $m_\chi$ is the DM mass. The sum is over the outer-shell $3p$ (16.08 eV binding energy) and $3s$ (34.76 eV binding energy) electrons. Figure~\ref{fig:ErDMshells} shows the contributions of the individual atomic shells to the total DM-electron scattering rate. For low electron recoil energies, the outer-shell contribution ($3p$) dominates, while at higher energy, the contribution from the $3s$ shell increases. This behavior becomes more pronounced as the DM mass increases. The same behavior is observed for the contribution from the $2p$ shell, although over the mass range considered here, contributions from the inner-shell orbitals are still negligible. This is in contrast to xenon, where contributions from the internal $n=4$ shell are significant. As a consequence, the expected ionization spectra in argon decrease more rapidly with recoil energy than for a xenon target.

\begin{figure}[]
\begin{center}
\includegraphics[width=\columnwidth]{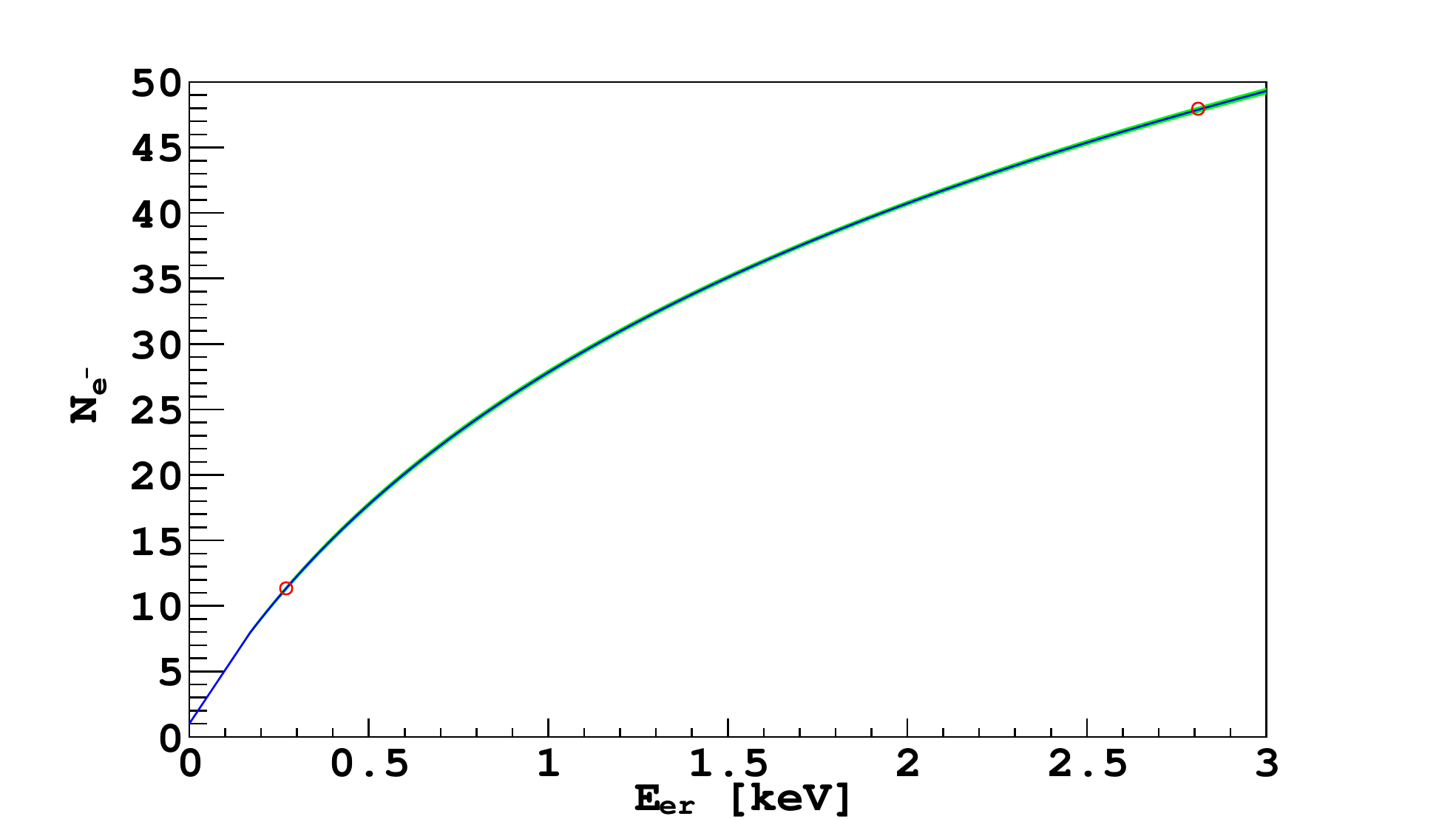}
\end{center}
\caption{\label{fig:ERScale}Calibration curve used to convert electron recoil spectra to ionization spectra. Below 8 $N_{e^-}$, we assume there is no recombination and use a straight line that intersects $N_{e^-}=1$ with a slope determined by the ratio of the number of excitations to ionization, $N_{ex}/N_i = 0.21$, measured in Ref.~\cite{Kubota:1976cf} and the work function measured in Ref.~\cite{doke:2002bo}. Above this point, the effects of recombination are included by fitting the Thomas-Imel model~\cite{Thomas:1987ek} to the mean N$_{e^-}$ measured for the \SI{2.82}{keV} K-shell and \SI{0.27}{keV} L-shell lines from the electron capture of $^{37}$Ar. In order to get good agreement between the model and data, we multiply the model by a scaling factor, whose best-fit value shifts the curve up by 15\%. This scaling factor can be interpreted as the agreement between our measured $N_{ex}/N_i$ and work function and the literature values. The green band shows the statistical uncertainty of the fit.}
\end{figure}

\begin{figure}[]
\begin{center}
\includegraphics[width=\columnwidth]{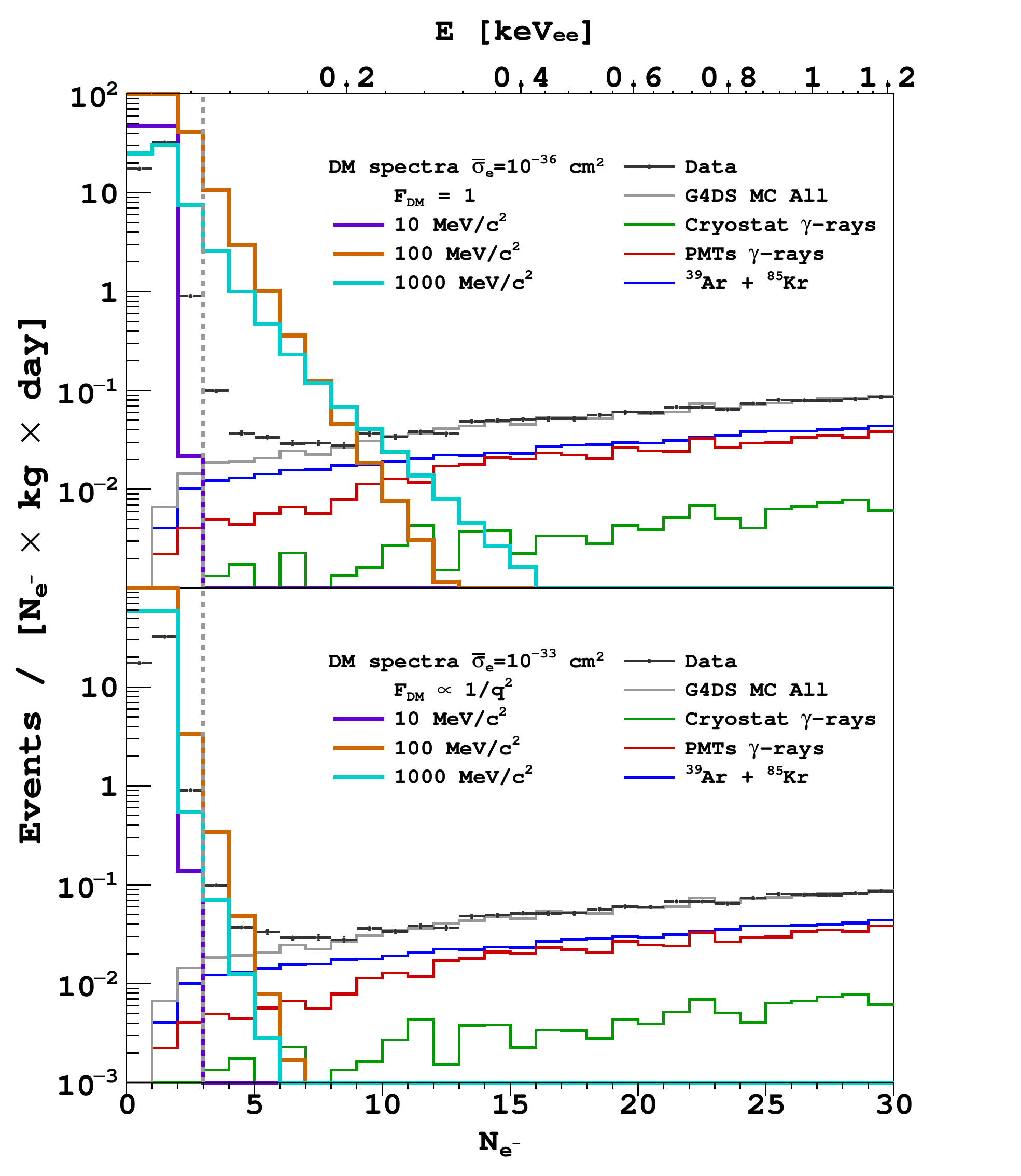}
\end{center}
\caption{\label{fig:NeSpecErDM_F1_MC}The 500 day \DSf\ ionization spectrum compared with predicted spectra from the \GFDS\  background simulation~\cite{Agnes:2017cz}. These are the same data and background spectra shown in Ref.~\cite{Agnes:2018vi}. Also shown are calculated DM-electron scattering spectra for DM particles with masses $m_\chi$ of \num{10}, \num{100}, and \SI{1000}{\MeV\per\square\c}, reference cross section $\sigmabar = $ \SI{E-36}{\square\cm} (top) and $\sigmabar = $ \SI{E-33}{\square\cm} (bottom), and \FDM $\, =1$ (top) and \FDM $\, \propto \,1/q^2$ (bottom). The vertical dashed line indicates the $N_{e^-}=3$ analysis threshold.}
\end{figure}

The calculated DM-electron recoil spectra are converted to the ionization spectra measured in DS-50 using a scale conversion based on a fit to low-energy peaks of known energy, as shown in Fig.~\ref{fig:ERScale} and described in Ref.~\cite{Agnes:2018vi}. The resulting ionization spectra are then smeared assuming the ionization yield and recombination processes follow a binomial distribution and convolved with the detector response, measured from single-electron events~\cite{Agnes:2018vi}. This procedure correctly reconstructs the measured width of the $^{37}$Ar K-shell (\SI{2.82}{keV}) and L-shell (\SI{0.27}{keV}) peaks. The expected DM-electron scattering ionization spectra in the case of a heavy mediator, $F_{\rm DM} = 1$, and in the case of a light mediator, $F_{\rm DM} \propto 1/q^2$, are shown in Fig.~\ref{fig:NeSpecErDM_F1_MC}.

We use a 500-day data set collected between 30~April, 2015, and 25~April, 2017, corresponding to a \SI{6786.0}{kg}{ d} exposure, to place limits on DM with masses below 1~GeV/c$^2$. The 500-day ionization spectrum used for the search is shown in Fig.~\ref{fig:NeSpecErDM_F1_MC}. Limits are calculated using a binned profile likelihood method implemented in RooStats~\cite{Cowan:2011cx,Moneta:2011jc,Verkerke:2006es}.  We use an analysis threshold of $\Ne = 3$, approximately equivalent to \SI{0.05}{keVee}, lower than the threshold used in Ref.~\cite{Agnes:2018vi}.  This increases the signal acceptance at the expense of a larger background rate from coincident single-electron events, which are not included in the background model and contribute as signal during the limit calculation.  The background model used in the analysis is determined by a detailed Monte Carlo simulation of the \DSf\ apparatus. Spectral features at high energy are used to constrain the simulated radiological activity within detector components to predict the background spectrum in the region of interest~\cite{Koh:2018}. The predicted spectrum is plotted alongside the data in Fig.~\ref{fig:NeSpecErDM_F1_MC} and described in greater detail in Ref.~\cite{Agnes:2018vi}. During the analysis, the overall normalization of the background model is constrained near its predicted value by a Gaussian nuisance term in the likelihood function. Additional Gaussian constraints on the background and signal spectral shape are included based on the uncertainty of the fit in Fig.~\ref{fig:ERScale} and the uncertainty in the S2-to-$\Ne$ conversion factor, extracted from single-electron data.

\begin{figure}[]
\begin{center}
\includegraphics[width=\columnwidth]{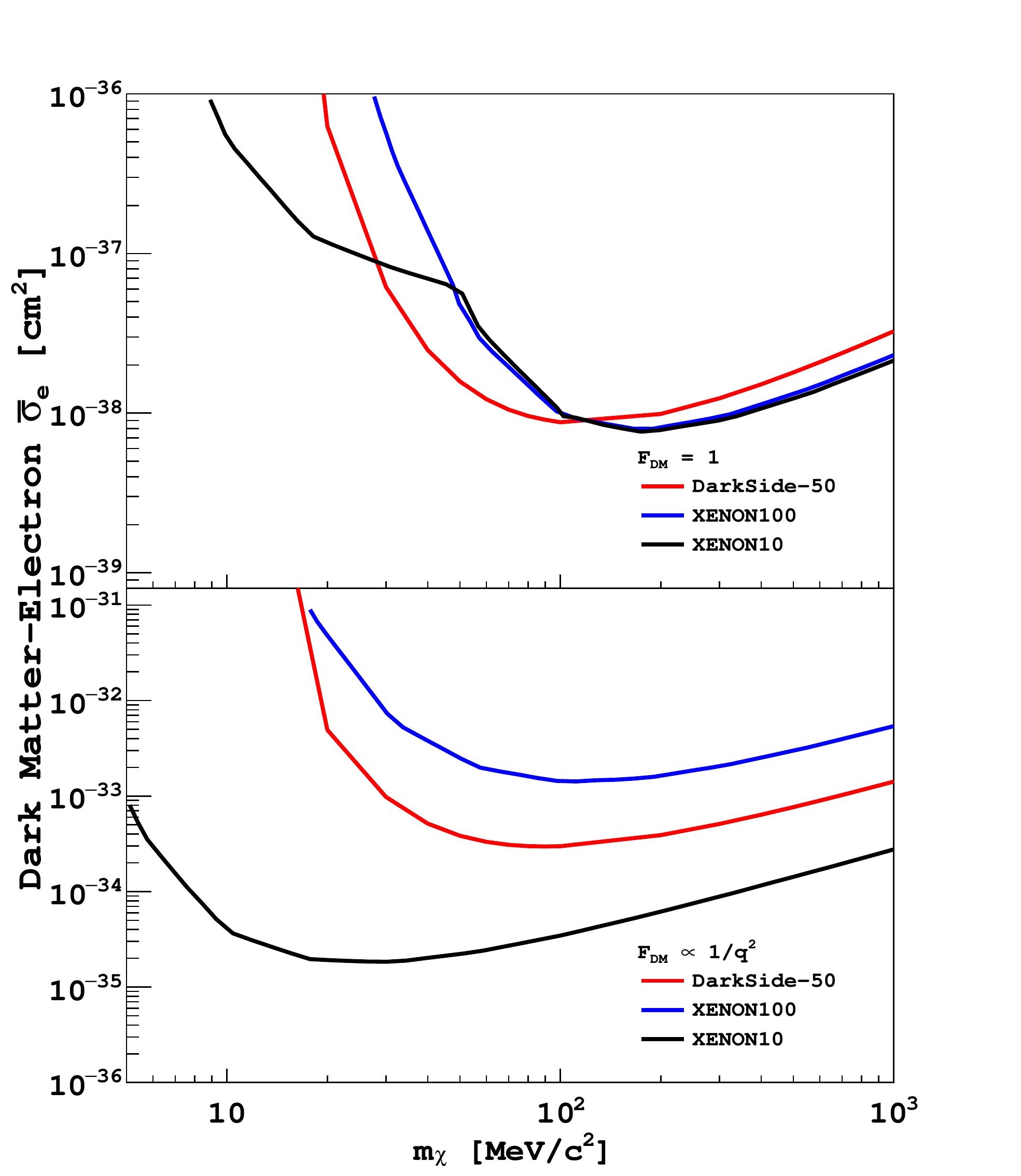}
\end{center}
\caption{\label{fig:Limits_ErDM_F1}\SI{90}{\percent} C.L. limits on the DM-electron scattering cross section for $F_{\rm DM} = 1$ (top) and $F_{\rm DM}\propto 1/q^{2}$ (bottom) for \DSf\ (red) alongside limits calculated in Ref.~\cite{Essig:2017bi} using data from XENON10 (black)~\cite{Angle:2011fl} and XENON100 (blue)~\cite{Aprile:2014he}.}
\end{figure}

The resulting 90\% C.L. limits are shown in Fig.~\ref{fig:Limits_ErDM_F1} for two assumptions of DM form factors, $F_{\rm DM}(q) = 1$ and $F_{\rm DM}(q) \propto {1/q}^2$. In the case of a light mediator, $F_{\rm DM}(q)\propto 1/q^{2}$, the constraints from DS-50 are not as stringent as the XENON10 experiment due to the higher ($N_{e^-}=3$) analysis threshold adopted in this Letter but are better than the XENON100 limit due the lower background rate. For a heavy mediator, $F_{\rm DM}(q)= 1$, we improve the existing limits from XENON10 and XENON100~\cite{Essig:2017bi} for dark-matter masses between 30~MeV/$c^2$ and 100~MeV/$c^2$, seeing a factor-of-3 improvement at 50~MeV/$c^2$.

\begin{acknowledgments}
The \DS\ Collaboration offers its profound gratitude to the \LNGS\ and its staff for their invaluable technical and logistical support.  We also thank the Fermilab Particle Physics, Scientific, and Core Computing Divisions.  Construction and operation of the \DSf\ detector was supported by the U.S. National Science Foundation (NSF) (Grants \grant{PHY}{0919363}, \grant{PHY}{1004072}, \grant{PHY}{1004054}, \grant{PHY}{1242585}, \grant{PHY}{1314483}, \grant{PHY}{1314501}, \grant{PHY}{1314507}, \grant{PHY}{1352795}, \grant{PHY}{1622415}, and associated collaborative grants \grant{PHY}{1211308} and \grant{PHY}{1455351}), the Italian Istituto Nazionale di Fisica Nucleare, the U.S. Department of Energy (Contracts \grant{DE}{FG02-91ER40671}, \grant{DE}{AC02-07CH11359}, and \grant{DE}{AC05-76RL01830}), the Russian Science Foundation (Grant \grant{16}{12-10369}), the Polish NCN (Grant \grant{UMO}{2014/15/B/ST2/02561}) and the Foundation for Polish Science (Grant \grant{Team2016}{2/17}).  We also acknowledge financial support from the French Institut National de Physique Nucl\'eaire et de Physique des Particules (IN2P3), from the UnivEarthS Labex program of Sorbonne Paris Cit\'e (Grants \grant{ANR}{10-LABX-0023} and \grant{ANR}{11-IDEX-0005-02}), and from the S\~ao Paulo Research Foundation (FAPESP) (Grant \grant{2016/09084}{0}).
\end{acknowledgments}

\bibliographystyle{ds}
\bibliography{ds,e_scattering}

\appendix*
\section{APPENDIX}

Here we provide additional details on the DM-electron scattering rate calculation described in the text. The explicit forms of the radial part of the wave function used to compute the atomic form factor, $\fsq$, are given by the Roothaan-Hartree-Fock  (RHF) wave functions~\cite{Bunge:1993km}, which are linear combinations of Slater-type orbitals:
\bea
R_{nl}(r) =&& a_0^{-3/2} \sum_{j} C_{jln} \frac{(2Z_{jl})^{n'_{jl}+ 1/2}}{\sqrt{(2n'_{jl})!}}\nonumber\\
&&\times\left(\frac{r}{a_0}\right)^{n'_{jl}-1}e^{-Z_{jl} r/a_0}\,,
\label{eq:RHFdef}
\eea
where the coefficients $C_{jln}$, $Z_{jl}$, and $n'_{jl}$ are given in Ref.~\cite{Bunge:1993km}.

In the literature, different procedures  have been used to approximate the outgoing electron wave function in such scattering scenarios.   One common approximation is to treat the final state as a pure plane-wave corrected by a Fermi factor,
\be
F(k', \Zeff) = \frac{2\pi \Zeff}{k'a_0}\frac{1}{1-e^{-2\pi \Zeff/(k'a_0)}} \,,
\ee
which parameterizes the distortion of the outgoing electron wave function by the effective screened Coulomb potential of the nucleus.  While the approximate shape of the ionization form factors, $f_\text{ion}^{nl}$, are consistent between the plane-wave solution and the continuum-state solution used in this work, the detailed structure does vary between the two.  At large momentum  transfers, the plane-wave and continuum solutions approach each other,  but they diverge at lower momentum transfers where the form factor is dominated by the overlap between the bound and continuum wave functions near the origin.   This is because the Fermi factor reproduces the behavior of the full wave function at the origin, but outer-shell orbitals have most of their support away from the origin, such that the overlap with the outgoing wave function is maximized away from the origin.  Thus, smaller atoms and inner shells have better agreement. For this reason, the discrepancy between using continuum versus plane-wave final states is smaller for argon than for xenon.  We, however, choose to use the full-continuum solutions for the presentation of all final results.  The continuum-state solutions to the Schr\"{o}dinger equation with potential $-\Zeff/r$ have radial wave functions indexed by $l$ and $k$, given by Ref.~\cite{Bethe:1977gr}
\bea
&&\widetilde{R}_{kl}(r) = (2\pi)^{3/2}(2kr)^l \frac{\sqrt{\frac{2}{\pi}}\left|\Gamma\left(l+1 - \frac{i\Zeff}{ka_0}\right)\right| e^{\frac{\pi \Zeff}{2ka_0}}}{(2l+1)!}\nonumber\\
&&\times e^{-ikr}\,{}_1F_1\left(l+1 + \frac{i\Zeff}{ka_0},2l+2, 2ikr\right).
\eea
The ratio of the wave function at the origin to the wave function at infinity gives the Fermi factor:
\be
\left | \frac{\widetilde{R}_{kl}(r=0)}{\widetilde{R}_{kl}(r=\infty)}\right|^2 = F(k,\Zeff)\,.
\ee
The normalization for these unbound wave functions is
\be
\int dr \, r^2 \, \tilde{R}^*_{kl}(r)\, \tilde{R}_{k'l'}(r) = (2\pi)^3 \frac{1}{k^2}\delta_{ll'} \delta(k-k')\,,
\ee
so that $\widetilde{R}_{kl}(r)$ itself is dimensionless. In terms of these wave functions, the ionization form factor is given by
\bea
&&\fsq = \frac{4k'^3}{(2\pi)^3}\sum_{l'} \sum_{L = |l'-l|}^{l'+l} {(2l+1)}(2l'+1)(2L+1)\nonumber\\
&&\times\left[  \begin{matrix} l & l' & L \\
      0 & 0 & 0 \\
   \end{matrix} \right]^2 \left |\int dr \, r^2 \widetilde{R}_{k'l'}(r) R_{nl}(r)j_L(qr)\right|^2\,
 \label{eq:fsqFull}
\eea
The term in brackets is the Wigner-$3j$ symbol evaluated at $m_1 = m_2 = m_3 = 0$, and $j_L$ is the spherical Bessel function of order $L$.

Following Refs.~\cite{Essig:2012it,Essig:2017bi}, the procedure used to determine $\Zeff$ is:
\begin{enumerate}
\item Treat the bound-state orbital $R_{nl}$ as a bound state of a pure Coulomb potential $-\Zeff^{nl}/r$, rather than the self-consistent potential giving rise to the RHF wave functions.
\item Determine $\Zeff^{nl}$ by matching the energy eigenvalue to the RHF eigenvalue.
\item Use this $\Zeff^{nl}$ to construct all $\widetilde{R}_{k'l'}(r)$ in the sum in Eq.~(\ref{eq:fsqFull}).
\end{enumerate}
For example, for the $3p$ shell of argon, $E_b^{3p} = 16.08~{\rm eV}$, so we solve
\[ 13.6~{\rm eV} \times \frac{(\Zeff^{3p})^2}{3^2} = 16.08~{\rm eV} \implies \Zeff^{3p} = 3.26\,.\]

\end{document}